\documentstyle[a4,12pt]{article}

\begin{document}

\newcommand{\zb}{\bar z}
\newcommand{\th}{\theta}
\newcommand{\tb}{\bar{\theta}}
\newcommand{\zt}{\tilde z}
\newcommand{\zbt}{\tilde{\zb}}
\newcommand{\tht}{\tilde{\th}}
\newcommand{\tbt}{\tilde{\tb}}
\newcommand{\TH}{\Theta}
\newcommand{\TB}{\bar{\Theta}}

\newcommand{\la}{\lambda}
\newcommand{\LA}{\Lambda}
\newcommand{\DE}{\Delta}
\newcommand{\OM}{\Omega}

\newcommand{\pa}{\partial}
\newcommand{\pab}{ \bar{\partial} }
\newcommand{\dab}{\bar D}

\newcommand{\pat}{ \tilde{\pa} }
\newcommand{\pabt}{ \tilde{\pab} }
\newcommand{\dt}{ \tilde{D} }
\newcommand{\dbt}{ \tilde{\dab} }

\newcommand{\ZB}{\bar Z}
\newcommand{\HT}{ {H_{\th}}^z }
\newcommand{\HB}{ {H_{\tb}}^z }
\newcommand{\HO}{ H_{\th} ^{\ \th} }
\newcommand{\HZ}{ H_{\zb} ^{\ \th} }
\newcommand{\HZB}{ H_{\zb} ^{\ z} }
\newcommand{\HOB}{ H_{\tb} ^{\ \th} }

\newcommand{\TT}{ {\cal T}_{\th z} }
\newcommand{\TZ}{ {\cal T}_{\tb \zb} }

\thispagestyle{empty}

\hfill    MPI-Ph/92-43
\vskip 0.01 truecm
\bigskip
\bigskip
\bigskip
\begin{center}
{\bf \Huge{Maurer-Cartan Forms and Equations}}
\end{center}
\begin{center}
{\bf \Huge{for}}
\end{center}
\begin{center}
{\bf \Huge{Two-Dimensional Superdiffeomorphisms}}
\end{center}
\vskip 0.8truecm
\bigskip
\centerline{{\bf Fran\c cois Gieres}$\, ^{\S} \, ^{\ddag}$}
\bigskip
\bigskip
\centerline{\it Max-Planck-Institut f\"ur Physik}
\centerline{\it Werner Heisenberg Institut}
\centerline{\it  F\"ohringer Ring 6}
\centerline{\it
D - 8000 - M\"unchen 40}
\vskip 3.7truecm

\nopagebreak
\begin{abstract}

We present
explicit expressions for the Maurer-Cartan forms
of the
superdiffeomorphism group associated to a super Riemann surface.
As an application to superconformal field theory, we use these
forms to evaluate the
effective action for the factorized
superdiffeomorphism anomaly.

\end{abstract}
\bigskip
\bigskip

\nopagebreak
\begin{flushleft}
\rule{2 in}{0.03cm} \\

{\footnotesize
\ ${}^{\S}$
Alexander von Humboldt Fellow.}
\\ [-0.04cm]
{\footnotesize
\ ${}^{\ddag}$
E-mail address: frg@dm0mpi11.}
 \\ [0.5cm]

MPI-Ph/92-43  \\
\vskip 0.07truecm
June 1992   \\
\end
{flushleft}

\newpage
\setcounter{page}{1}

\section{Introduction}

Maurer-Cartan (MC) forms and equations are familiar to
mathematicians and physicists
alike: they are used for instance to define connections
on homogenous spaces \cite{kn} or to construct field theories
on group manifolds \cite{tr}.
While the case of finite-dimensional Lie groups is well known,
the infinite-dimensional theory is less familiar.
An example for the latter is provided by the group of
diffeomorphisms of a smooth manifold \cite{jm}.
The particular case of $C^{\infty}$-diffeomorphisms on a Riemann surface
has recently been investigated \cite{sl}. The corresponding
results have been
applied to evaluate the Wess-Zumino (WZ) action associated to
the chirally split diffeomorphism anomaly on the complex plane
\cite{sl}; this anomaly occurs in conformal models and is equivalent
to the Weyl anomaly \cite{klt}.
It is quite remarkable that
the WZ functional can be explicitly determined in this case
despite the fact that
the symmetry group is non-Abelian. (E.g., for non-Abelian gauge
theories
in four dimensions, a complete evaluation of the WZ action has not been
achieved to date.) The computational
success in the two-dimensional case ultimately relies
on the low dimensionality of
the underlying space-time manifold.

Some time ago,
the MC-forms and equations have been generalized
to finite-dimensional supergroups,
e.g. to the graded Poincar\'e group
\cite{nr} \cite{tr}. In the present paper, we consider the
infinite-dimensional case and, more specifically, the superdiffeomorphism
group associated with a $N=1$ super Riemann surface (SRS). Along the lines
of the bosonic theory \cite{sl}, we determine the corresponding MC-forms
and we use them to construct the WZ action associated to the chirally
split superdiffeomorphism anomaly on the complex superplane.
This anomaly has been known for some time \cite{dg}
and the explicit expression for the associated
WZ action has previously been
postulated in analogy to the results of the non-supersymmetric
theory \cite{dg1}.

As to the organization of the paper, we first discuss the case
of generic SRS's and subsequently consider the limitation of
supercomplex structures which is frequently chosen for SRS's
(section 3).

\section{General case}

\subsection{Supercomplex structures}

Let us first recall the basic facts which are needed in the sequel
(see \cite{dg} and references therein).
A SRS is locally parametrized by coordinates
($Z, \ZB , \TH , \TB$). The
super 1-forms
$e^Z \equiv dZ + \TH \, d\TH \, , \, e^{\TH} \equiv d\TH$ (and c.c.)
which span the cotangent space to the SRS
can be expressed with respect to a
reference system
$e^z \equiv dz + \th \, d\th \, , \, e^{\th} \equiv d\th$ (and c.c.) by
\begin{eqnarray}
e^Z & = & \left[ \, e^z \ + \ e^{\zb} \, \HZB \ + \ e^{\th} \, \HT \ + \
e^{\tb} \, \HB \, \right] \, \LA
\label{30} \\
e^{\TH} & = &
\left[ \, e^z \ + \ e^{\zb} \, \HZB \ + \ e^{\th} \, \HT \ + \
e^{\tb} \, \HB \, \right] \, \tau \ + \
\left[ \, e^{\th} \, \HO  \ + \ e^{\zb} \, \HZ \ + \
e^{\tb} \, \HOB \, \right] \, \sqrt{\LA}
\nonumber
\end{eqnarray}
(and the complex conjugate expressions). The structure equations
$de^Z + e^{\TH} \, e^{\TH} \, = \, 0 \, = \, de^{\TH}$ (and c.c.)
imply that the odd superfields $\HB , \HT$ (and c.c.) are the only
independent Beltrami coefficients and that all the others depend
on them and their derivatives.
Furthermore, these equations imply that the
factor $\tau$ depends on the integrating factor $\LA$
and that $\LA$
satisfies the linear differential equation
\begin{equation}
\label{39}
\tilde{\bar{D}}\, {\rm ln} \, \LA  \ = \
 \pa \HB \, - \, \frac{\HOB}{\HO} \, \pa \HT
\ \ \ .
\end{equation}
Here, $\tilde{\dab}$ is a
linear differential operator whose coefficients are functions of
$\HB, \HT$ and their derivatives;
the canonical derivations in superspace are denoted by
\[
\pa \, \equiv \,
\frac{\pa}{\pa z}
\ \ \ \ , \ \ \ \
D\, \equiv \,
\frac{\pa}{\pa \th} \, + \, \th \,
\frac{\pa}{\pa z}
\ \ \ \ \ \ , \ \ \ \ \ \
( \ D^2 \ = \
\pa \ )
\ \ \ \ \ \ \ \ \
{\rm and \ \, c.c.}
\ \ \ .
\]

The superfield $\HB$ admits a component field expansion
\begin{equation}
\label{33}
\HB \ = \ \sigma_{\tb} ^{\ z} \ + \ \th \, v^z \ + \ \tb \,
\mu_{\zb} ^{\ z} \ + \ \th \tb \, [-i \alpha_{\zb} ^{\ \th} ]
\ \ \ ,
\end{equation}
where
$\mu_{\zb} ^{\ z} (z, \zb )$ and $
\alpha_{\zb} ^{\ \th} (z, \zb )
$ denote the usual Beltrami coefficient and its
fermionic partner, respectively.

Infinitesimal
superdiffeomorphisms generated by the vector field
\[
\Xi \cdot \pa \ \equiv \
\Xi^{z} (z , \zb , \th , \tb )\, \pa \ + \
\Xi^{\zb} (z , \zb , \th , \tb )\, \pab  \ + \
\Xi^{\th} (z , \zb , \th , \tb )\, D \ + \
\Xi^{\tb} (z , \zb , \th , \tb )\, \bar{D}
\]
act on the basic variables according to
\begin{eqnarray}
s \, \TH & = &
i_{\Xi \cdot \pa} \, d\TH \ = \
i_{\Xi \cdot \pa} \,  e^{\TH}
\ = \
C^z \, \tau \, + \, C^{\th} \, \sqrt{\LA}
\nonumber   \\
s \, Z & = &
i_{\Xi \cdot \pa} \, dZ \ = \
i_{\Xi \cdot \pa} \, \left[ \, e^Z \, - \, \TH \, e^{\TH} \, \right]
\ = \
C^z \, \LA \, -\, \TH \,
\, (s \, \TH )
\ \ \ .
\label{312}
\end{eqnarray}
Here, $i_{\Xi \cdot \pa}$ denotes the interior product with the vector field
$\Xi \cdot \pa$ and the parameters
\begin{eqnarray}
\label{14a}
C^z & \equiv & \Xi^z \ + \ \Xi^{\zb} \, H_{\zb} ^{\ z} \ + \
\Xi ^{\th} \, \HT \ + \ \Xi ^{\tb} \, \HB
\\
C^{\th} & \equiv &
\Xi ^{\th} \, \HO \ + \ \Xi ^{\zb} \, \HZ \ + \
\Xi ^{\tb} \, H_{\tb} ^{\ \th}
\nonumber
\end{eqnarray}
are supposed to
represent ghost superfields; the nilpotency requirement
for the BRS-operation $s$ then leads to the
transformation laws
\begin{eqnarray}
\label{444}
s C^z & = & -\, \left[ \, C^z \, \pa C^z \ + \ C^{\th} \,
C^{\th} \, \right]
\\
s C^{\th} & = & -\, \left[ \, C^z \, \pa  C^{\th} \ + \  \frac{1}{2}
\, C^{\th} \, (\pa C^z ) \, \right]
\ \ \ .
\nonumber
\end{eqnarray}
The induced variations of the independent Beltrami fields and of the
integrating factor read
\begin{eqnarray}
s \HB & = &  (\, \bar{D} \ - \ \HB \, \pa \, ) \, C^z
\ + \  ( \pa \HB ) \, C^z \ - \ 2 \,
{H_{\tb}}^{\th} \, C^{\th}
\nonumber  \\
s \HT & = & (\, D  \ - \ \HT \, \pa  \, ) \, C^z
\ + \  ( \pa \HT ) \, C^z \ - \ 2 \, \HO
\, C^{\th}
\nonumber   \\
s \LA & = & \pa \, (C^z \, \LA)  \ + \ 2 \, C^{\th} \, \tau \,
\sqrt{\LA}
\ \ \ .
\label{14}
\end{eqnarray}

\subsection{Maurer-Cartan form for the
superdiffeomorphism group}

In this section, we
determine the MC 1-form associated to the supergroup
${\rm Diff}_0 ({\bf S\Sigma})$, i.e. the
superdiffeomorphisms on a (compact) SRS ${\bf S\Sigma}$
which are homotopic
to the identity.

By virtue of eqs.(\ref{312}),
the infinitesimal superdiffeomorphisms act on the local
coordinates ($Z, \ZB, \TH , \TB$)
according to
\begin{eqnarray}
sZ & = & C^z \, \LA \ - \ \TH \, s\TH
\ \ \ \ \ \ \ \ \ \ {\rm and \ c.c.}
\label{70}
\\
s\TH & = & C^z \, \tau \ + \ C^{\th} \, \sqrt{\LA}
\ \ \ \ \ \ \ \ {\rm and \ c.c.}
\ \ \ .
\nonumber
\end{eqnarray}
These relations can be `solved' for the ghost fields $C^z$ and $C^{\th}$,
respectively :
\begin{eqnarray}
C^z & = & \LA^{-1} \ \left[ \, sZ \ + \ \TH \, s\TH \, \right]
\ \ \ \ \ \ \ \ \ \ \ \ \ \ \ \ \ \ \ \ \ \ \ {\rm and \ c.c.}
\label{71}
\\
C^{\th} & = & \LA^{-3/2} \ \left[ \, (\LA +  \TH \tau ) \, s\TH
\ - \ \tau \, sZ \, \right]
\ \ \ \ \ \ \ \ \ {\rm and \ c.c.}
\ \ \ .
\nonumber
\end{eqnarray}
The integrating factors are locally given by
$\LA = \pa Z  + \TH \, \pa \TH \, , \, \tau = \pa \TH$
and the induced variations of $C^z \, , \, C^{\th}$ are those given by
eqs.(\ref{444}).

{}From the previous considerations, we can deduce the components
$\OM^z \, , \, \OM^{\th}$ (and c.c.) of the {\em Maurer-Cartan
1-form} associated to the group ${\rm Diff}_0 ({\bf S\Sigma})$.
To do so, we rely on the fact \cite{rs} \cite{sl} that there is a canonical
morphism between the BRS algebra (with differential $s$ and generators
$C^z , C^{\zb} , C^{\th} , C^{\tb}$) and the algebra of differential forms
on
${\rm Diff}_0 ({\bf S\Sigma})$ (with differential $\delta$ and generators
$\OM^z ,\OM^{\zb}, \OM^{\th} ,\OM^{\tb}$).
This morphism is realized by
taking
eqs.(\ref{71}) and
substituting $s$ by $\delta$ and $Z, \TH$ by $Z^{\varphi} ,
\TH^{\varphi}$ with $\varphi \in {\rm Diff}_0 ({\bf S \Sigma})$ :
\begin{eqnarray}
\OM^z &=& (\pa Z^{\varphi} \, + \, \TH^{\varphi}
\pa \TH^{\varphi} )^{-1} \ \left[ \, \delta Z^{\varphi}
\ + \ \TH^{\varphi} \, \delta \TH^{\varphi} \, \right]  \
\ \ \ \ \ \
\ \ \ \ \ \ \ \ \ \ \ \ \ \ \ \ \ \ \ \ \ \ \ \ \ \ \ {\rm and \ c.c.}
\label{72}
\\
\OM^{\th} & = & (\pa Z^{\varphi} \, + \, \TH^{\varphi}
\pa \TH^{\varphi} )^{-3/2} \ \left[ \,
(\pa Z^{\varphi} \, + \, 2 \TH^{\varphi} \pa \TH^{\varphi}
 ) \, \delta \TH^{\varphi} \ - \
( \pa \TH^{\varphi} ) \, \delta Z^{\varphi} \, \right]
\ \ \ \   {\rm and \ c.c.}
\ .
\nonumber
\end{eqnarray}
By construction (cf.eqs.(\ref{444})),
these 1-forms satisfy the {\em Maurer-Cartan equations}
\begin{eqnarray}
\delta \OM^z & = & -\, \left[ \, \OM^z \, \pa \OM^z \ + \ \OM^{\th} \,
\OM^{\th} \, \right]
\ \ \ \ \ \ \ \ \ \ \ \ \ \ \ \ {\rm and \ c.c.}
\label{73}
\\
\delta \OM^{\th} & = & -\, \left[ \, \OM^z \, \pa  \OM^{\th} \ + \  \frac{1}{2}
\, \OM^{\th} \, (\pa \OM^z ) \, \right]
\, \ \ \ \ \ \ \ \   {\rm and \ c.c.}
\ \ \ .
\nonumber
\end{eqnarray}
For further mathematical details, we refer to \cite{sl}.

\subsection{Wess-Zumino action}

Let $\int_{{\bf S\Sigma}} d^4 z\;
{\cal A} (C^z , C^{\th} \, ; \, \HB , \HT ) \ + \ {\rm c.c.}$
denote the
{\em chirally split superdiffeomorphism anomaly} on ${\bf S\Sigma}$.
According to the algebraic treatment of anomalies \cite{rs},
the {\em WZ action} associated to this
anomaly is given by
\begin{equation}
\label{74}
\Gamma_{WZ} [ \varphi \, ; \HB , \HT ] \ + \ {\rm c.c.} \ = \
- \, \int_{0}^{1} \int_{{\bf S\Sigma}} d^4 z \
{\cal A} \left( \, \OM_t ^z , \OM_t ^{\th} \, ;  \,
(\HB)^{\varphi_t} , (\HT)^{\varphi_t} \, \right)
\ + \ {\rm c.c.}
\ \ \ ,
\end{equation}
where
$\varphi \in {\rm Diff}_0 ({\bf S \Sigma})$ transforms like
$s (\HB )^{\varphi} = 0$ or, equivalently,
$s \varphi = \Xi \cdot \pa \, \varphi$ \cite{rs} \cite{sl} and where one has
by construction
\[
s \Gamma_{WZ} [ \varphi \, ; \HB , \HT ]  \ = \
\int_{0}^{1} \int_{{\bf S\Sigma}} d^4 z \
{\cal A} (C^z , C^{\th} \, ; \, \HB , \HT )
\ \ \ .
\]
In the expression (\ref{74}),
$\varphi_t$ denotes a smooth family of superdiffeomorphisms
which interpolates between the identity and
$\varphi$.
The 1-form $\OM_t$ is obtained from the MC 1-form $\OM$ by
replacing $\delta$ by $d_t \equiv dt\, \pa /\pa t$
and $\varphi$ by $\varphi_t$.
The `$H^{\varphi_t}$' follow from the `$H$' by the action of the finite
supercoordinate transformation $\varphi_t$ \cite{dg}.

The integral (\ref{74}) is rather complicated, even if we
consider the superplane,
${\bf S\Sigma} = {\bf SC}$, and if we restrict the geometry by
$\HT = 0 = H_{\tb} ^{\ \zb}$ (see next section): in this case,
the factorized superdiffeomorphism
anomaly
is explicitly given by \cite{dg}
\begin{equation}
\int_{{\bf SC}} d^4 z\;
{\cal A} (C^z ; \HB )
\ + \ {\rm c.c.} \ = \
\int_{{\bf SC}} d^4 z\; C^z \, \pa^2 D \HB
\ + \ {\rm c.c.}
\ \ \ .
\label{313}
\end{equation}
In section 3.2,
we will work out the expression (\ref{74}) for this case.

\section{Restriction of the geometry ($\HT = 0$)}

\subsection{Geometric framework}

For $\HT = 0$ and
\[
0 \ = \ s\HT \ = \ D C^z \ - \ 2 \, C^{\th}
\ \ \ ,
\]
the $s$-variations of the basic variables reduce to
\begin{eqnarray}
s \HB & = &  \left[ \, \dab \ - \ \HB \, \pa
\ + \ \frac{1}{2} \, (D \HB ) D \, \right] \, C^z
\ + \  ( \pa \HB ) \, C^z
\label{16a} \\
s C^z & = & -\, \left[ \, C^z \, \pa C^z \ + \ \frac{1}{4} \,
(D C^z )^2  \, \right]
\nonumber \\
s\LA & = & C^z \, \pa \LA \ + \ \frac{1}{2} \, (DC^z ) \, D \LA
\ + \ \LA \, \pa C^z
\ \ \ .
\nonumber
\end{eqnarray}
Moreover, the integrating factor equation, eq.(\ref{39}),
takes the simple form
\begin{equation}
\label{75}
\left[ \, \dab \ - \ \HB \, \pa \ + \ \frac{1}{2} \, (D\HB ) \, D \, \right]
\, \LA \ = \ (\pa \HB ) \, \LA
\ \ \ .
\end{equation}

{}From the local form of $\HT$ \cite{dg}, i.e.
$\HT = (\pa Z + \TH \pa \TH )^{-1} (DZ - \TH D\TH )$, and the condition
$\HT = 0$,
we deduce
\begin{eqnarray}
DZ & = & \TH \, D\TH
\nonumber \\
(D\TH )^2 & = &
\pa Z + \TH \, \pa \TH
\ \ \ ,
\label{76}
\end{eqnarray}
where the second equation follows from the first one by application of $D$.
It should be emphasized that $Z$ and $\TH$ still depend on both $z , \th$ and
$\zb , \tb$.

For consistency, we only consider those superdiffeomorphisms
\[
\left( z , \zb , \th , \tb \right) \longrightarrow
\left( \, z^{\prime} (z , \zb , \th , \tb ) \, , \,
\zb^{\prime} (z , \zb , \th , \tb ) \, , \,
\th^{\prime} (z , \zb , \th , \tb ) \, ,   \,
\tb^{\prime} (z , \zb , \th , \tb ) \, \right)
\]
which respect the condition $\HT=0$, i.e. satisfy \cite{dg}
\begin{equation}
\label{78}
0 \ = \ [\, Dz^{\prime} \, - \, \th^{\prime} \, D \th^{\prime} \, ] \ + \
[\, D\zb^{\prime} \, - \, \tb^{\prime} \, D \tb^{\prime} \, ] \;
H_{\zb^{\prime}}^{\ z^{\prime}} \ + \
[\, D \tb^{\prime}  \, ] \;
H_{\tb^{\prime}}^{\ z^{\prime}}
\ \ \ .
\end{equation}

\subsection{Construction of the WZ action on ${\bf SC}$}

For $\HT =0$, the
MC equations, eqs.(\ref{73}), reduce to
\begin{equation}
\delta \OM^z \ = \ -\, \left[ \, \OM^z \, \pa \OM^z \ + \
\frac{1}{4} \; (D\OM ^z )\,(D\OM ^z ) \, \right]
\ \ \ \ \ \ \ \ \ \ \ \ \ \ {\rm and \ c.c.}
\ \ \ ,
\label{73a}
\end{equation}
with
\[
\OM^z \ = \ \frac{\delta Z^{\varphi} \, + \,
\TH^{\varphi} \, \delta \TH^{\varphi}}
{\pa Z^{\varphi} \, + \, \TH^{\varphi}
\pa \TH^{\varphi} }
\ = \
\frac{\delta Z^{\varphi}
\, - \, (\delta \TH^{\varphi} ) \,  \TH^{\varphi} }
{(D \TH^{\varphi} )^2 }
\ \ \ .
\]
Furthermore, the WZ action, eq.(\ref{74}), becomes
\begin{equation}
\label{74a}
\Gamma_{WZ} [ \varphi \, ; \HB ] \ + \ {\rm c.c.} \ = \
- \, \int_{0}^{1} \int_{{\bf SC}} d^4 z \
{\cal A} \left( \, \OM_t ^z \, ; \, (\HB)^{\varphi_t}  \, \right)
\ + \ {\rm c.c.}
\ \ \ .
\end{equation}
Here, ${\cal A}$ is given by eq.(\ref{313}) and
\begin{equation}
\label{308}
(\HB)^{\varphi} \ = \
\frac{ \dab Z^{\varphi} - \TH^{\varphi} \,
\dab \TH^{\varphi} }{
\pa Z^{\varphi} + \TH^{\varphi} \,
\pa \TH^{\varphi} }
\ = \
\frac{ \dab Z^{\varphi} - (\dab \TH^{\varphi} ) \,
\TH^{\varphi} }{
(D \TH^{\varphi} )^2 }
\ \ \ ,
\end{equation}
where the superdiffeomorphism
$\varphi
(z , \zb , \th , \tb) \, \equiv \,
(z^{\prime} , \zb^{\prime} , \th^{\prime} , \tb^{\prime})$
is subject to the condition (\ref{78}).
The evaluation of the functional (\ref{74a})
is fairly lengthy; roughly, it proceeds along
the lines of the bosonic theory \cite{sl}, though
some new arguments have to be invoked due to the anticommuting variables.
Before discussing the calculation, we state the
{\em final result}:
\begin{equation}
\label{74b}
-2 \, \Gamma_{WZ} [ \varphi \, ; \HB ]  \ = \
\int_{{\bf SC}} d^4 z \
(\HB)^{\varphi} \, \pa D \, {\rm ln} \, \LA^{\varphi} \ - \
\int_{{\bf SC}} d^4 z \
\HB \, \pa D \, {\rm ln} \, \LA
\end{equation}
with $\LA \, \equiv \, \pa Z + \TH \pa \TH \, = \, (D\TH )^2$.

Let us now
sketch the main steps of the derivation.
By virtue of eq.(\ref{313}),
\begin{eqnarray}
\label{300}
\int_{{\bf SC}} d^4 z \
{\cal A} \left( \, \OM_t ^z \, ; \, (\HB)^{\varphi_t}  \, \right)
& = &
\int_{{\bf SC}} d^4 z \
\OM_t^z \ \pa^2 D
(\HB)^{\varphi_t}
\\
& = & \frac{1}{2} \,
\int_{{\bf SC}} d^4 z \; \left\{ \,
\OM_t^z \ \pa^2 D
(\HB)^{\varphi_t}
\ + \
(\HB)^{\varphi_t} \ \pa^2 D
\OM_t^z \, \right\}
\ \ \ ,
\nonumber
\end{eqnarray}
henceforth we
consider the symmetrized integrand
\begin{equation}
\label{301}
{\cal A} \left( \, \OM \, ; \HB  \, \right)
\ = \
\frac{1}{2} \, \left[ \,
\frac{ \delta Z - (\delta \TH ) \, \TH}{(D\TH )^2} \ \pa^2 D \
\frac{ \dab Z - (\dab \TH ) \, \TH}{(D\TH )^2}
\ + \ ( \ \delta \leftrightarrow \dab \ ) \, \right]
\ \ \ .
\end{equation}
Here and in the following, we have suppressed the suffixes $t$ and $\varphi$
in order to simplify the notation.
Note that $\delta$ and $\dab$ cannot simply be exchanged, because
the first is an
even and the second an odd operator (e.g. $\delta D = D \delta$, but
$\dab D = - \dab D$); thus, the two contributions in eq.(\ref{301})
have to be evaluated separately.
Our aim is to show that
\begin{equation}
\label{302}
{\cal A} \left( \, \OM ; \HB  \, \right)
\ = \
\delta \, \left[ \, \HB
\, \pa D \, {\rm ln} \, D\TH \, \right]
\ + \ \pa \, [...]
\ + \ \pab \, [...]
\ + \ D \, [...]
\ + \ \dab \, [...]
\ \ \ ,
\end{equation}
which relation immediately yields
the result (\ref{74b}) by integration over the suppressed variable $t$.

First, we evaluate
$\pa^2 D \left[ (D\TH )^{-2} ( \dab Z - (\dab \TH ) \, \TH ) \right]$
and express all derivatives $D Z$ and $\pa Z$ in terms of $D\TH$ and
$\pa \TH$ by virtue of eqs.(\ref{76}).
For the second term in eq.(\ref{301}), we proceed in a similar way.
Thus, one is led to
\begin{eqnarray}
\label{303}
- \, {\cal A} \left( \, \OM \, ; \HB  \, \right)
& = &
\frac{ \delta Z - (\delta \TH ) \, \TH}{(D\TH )^2} \
\frac{ \dab (\pa^2 \TH )}{D\TH}
\ - \
\frac{ \dab Z - (\dab \TH ) \, \TH}{(D\TH )^2} \
\frac{ \delta (\pa^2 \TH )}{D\TH}
\\
& &
\ + \ {\rm terms \ \, in} \
(D\TH )^{-4} ,...,
(D\TH )^{-7}
\ \ \ .
\nonumber
\end{eqnarray}
According to the Leibniz rule,
the first term on the r.h.s. can be rewritten as
\[
\dab \left[ \frac{ \delta Z - (\delta \TH ) \, \TH}{(D\TH )^2} \;
\frac{\pa^2 \TH }{D\TH} \right] \ - \
\dab \left[ \frac{ \delta Z - (\delta \TH ) \, \TH}{(D\TH )^3} \right]
\pa^2 \TH
\ \ \ .
\]
In this expression, we replace $\pa^2 \TH \, / \, D\TH$
by the super Schwarzian derivative \cite{df}:
\begin{eqnarray}
{\cal S} (Z , \TH \, ; z , \th ) & \equiv &
\pa D \, {\rm ln} \, D \TH \ - \ ( D \, {\rm ln} \, D \TH ) \,
\ (\pa \, {\rm ln} \, D \TH )
\label{304}
\\
& = &
\frac{\pa^2 \TH }{D\TH} \ - \ 2 \
\frac{(\pa \TH )(\pa D \TH )}{(D\TH )^2}
\ \ \ .
\nonumber
\end{eqnarray}
Proceeding in this way, one obtains a formula for $-{\cal A}$
which involves a lengthy expression; however the latter can be
rewritten in a compact way
by virtue of the following useful equations:
\[
-2
\frac{ \delta Z - (\delta \TH )  \TH}{(D\TH )^2} \;
\frac{ \dab Z - (\dab \TH ) \TH }{(D\TH )^2}
\left[
\frac{\pa^3 \TH }{D\TH}  -  3
\frac{(\pa^2 D \TH )(\pa \TH ) }{(D\TH)^2}  -  6
\frac{(\pa D \TH )(\pa^2 \TH ) }{(D\TH)^2}    +    12
\frac{(\pa D \TH )^2 (\pa \TH ) }{(D\TH)^3}    \right]
\]
\begin{eqnarray}
& = &
-2 \,
\left[ \, \delta Z - (\delta \TH ) \, \TH  \, \right] \,
\left[ \, \dab Z - (\dab \TH ) \, \TH  \, \right]
\, D \, \left\{ \,
\frac{1}{D\TH} \, D \, \left[ \,
\frac{{\cal S}}{(D\TH)^3} \, \right] \, \right\}
\label{305}
\\
& =  &
4 \;
\frac{ \delta Z - (\delta \TH )  \TH}{(D\TH )^2} \;
\frac{ \dab D \TH }{D\TH} \ {\cal S}
\ + \ 4 \;
\frac{ \dab Z - (\dab \TH ) \TH }{(D\TH )^2}
\ \frac{ \delta D \TH }{D\TH}  \ {\cal S}
\ + \ D \, [ ...]
\nonumber
\ \ \ .
\end{eqnarray}
Here, we introduced the notation ${\cal S} \equiv
{\cal S} (Z , \TH \, ; z , \th )$ and we repeatly used the Leibniz rule and
eqs.(\ref{76}) to pass to the last line. In conclusion,
one finds the intermediate result
\begin{eqnarray}
- {\cal A} \left( \, \OM \, ; \HB  \, \right)
& = &
\dab \, \left[ \, \frac{ \delta Z - (\delta \TH ) \, \TH}{(D\TH )^2} \
{\cal S} \, \right]
\ - \
\delta \, \left[ \, \frac{ \dab Z - (\dab \TH ) \, \TH}{(D\TH )^2} \
{\cal S} \, \right]
\nonumber
\\
& &
+ \ 2 \
\frac{ \dab Z - (\dab \TH ) \, \TH}{(D\TH )^2} \
\frac{\delta D \TH}{D \TH} \
{\cal S}
\ - \ 2 \
\frac{ \delta Z - (\delta \TH ) \, \TH}{(D\TH )^2} \
\frac{\dab D \TH}{D \TH} \
{\cal S}
\label{306}
\\
& &
+  \ 2 \
\frac{ (\delta \TH )( \dab \TH )}{(D\TH )^2} \
{\cal S}
\ + \  D [ ...]
\nonumber
\ \ \ .
\end{eqnarray}
Next, we consider the contribution
\begin{equation}
\label{307}
2 \
\frac{ \dab Z - (\dab \TH ) \, \TH}{(D\TH )^2} \
\frac{\delta D \TH}{D \TH} \
{\cal S}
\ =\ 2\, \HB \, (\delta \, {\rm ln} \, D\TH ) \, {\cal S}
\ \ \ .
\end{equation}
In order to evaluate the r.h.s., we apply $\delta$ to the Beltrami equation
(\ref{308}), i.e. to
\[
\dab Z \, - \, (\dab \TH ) \, \TH \, - \, \HB \, ( D\TH )^2  \ =\ 0
\ \ \ ,
\]
and obtain the relation
\begin{equation}
\label{309}
- 2\, \HB \, (\delta \, {\rm ln} \, D\TH ) \ = \  \delta \HB \ - \
\frac{ \delta [ \dab Z - (\dab \TH ) \, \TH ]}{(D\TH )^2}
\ \ \ .
\end{equation}
Substituting this result and eq.(\ref{304}) into the expression
(\ref{307}), we get
\[
2\, \HB \, (\delta \, {\rm ln} \, D\TH )  \, {\cal S}
\ = \  - \, (\delta \HB ) \, \pa D \, {\rm ln} \, D\TH \ + \
(\delta \HB ) \, (\pa  \, {\rm ln} \, D\TH ) \,
( D  \, {\rm ln} \, D\TH ) \ + \ ...
\ \ \ .
\]
A convenient form for the second term on the r.h.s. follows by
application of $\delta$ to the integrating factor equation
(\ref{75}):
\begin{equation}
\label{310}
(\delta \HB ) \, \pa  \, {\rm ln} \, D\TH  \ = \
- \frac{1}{2} \, \pa \delta \HB \; + \;
\frac{1}{2} \, ( D \delta \HB ) \,
( D \, {\rm ln} \, D\TH )
\; + \;
\left[ \dab - \HB \pa +
\frac{1}{2} \, ( D \HB ) D \right] \,
\delta  \, {\rm ln} \, D\TH
{}.
\end{equation}
By substituting the previous expressions into
eq.(\ref{306}), one is led to the advocated equation
(\ref{302}) and thereby to the final result (\ref{74b}).

\section{Concluding remarks}

As noted in reference \cite{dg1},
the superdiffeomorphism
$\varphi$ can be further restricted by the condition $
(\HB)^{\varphi} =0$. Then, the WZ action
(\ref{74b}) reduces to the so-called \cite{sl}
{\em Wess-Zumino-Polyakov functional},
\begin{equation}
\Gamma_{WZP} [ \HB ] \ = \
\frac{1}{2} \, \int_{{\bf SC}}
d^4 z \
\HB \, \pa  D  \, {\rm ln} \, \LA
\label{79}
\ \ \ .
\end{equation}
This expression represents a very compact notation for the
superspace generalization of Polyakov's chiral gauge action
\cite{gn} \cite{dg1}.

We note that the case of $(p,q)$ supersymmetry can be treated
along the lines of our previous discussion. As an application,
we explicitly constructed
the WZ action associated to the superdiffeomorphism anomaly
$\int_{{\bf SC}} d^3 z \, C^z \, \pa^2 D \HZB$
\cite{gg} \cite{dg}
in the $z$-sector of the $(1,0)$ supersymmetric theory.
In this case, one has to evaluate the expression
\begin{equation}
{\cal A}^{(1,0)} \left( \, \OM \, ; \HZB  \, \right)
\ = \
\frac{1}{2} \, \left[ \,
\frac{ \delta Z + \TH \, (\delta \TH )}{(D\TH )^2} \ \pa^2 D \
\frac{ \pab Z + \TH \, (\pab \TH )}{(D\TH )^2}
\ - \ ( \ \delta \leftrightarrow \pab \ ) \, \right]
\ \ \ .
\end{equation}
Since both $\delta$ and $\pab$ are even, the second contribution
simply follows by antisymmetrizing the result of the first one.
{}From the intermediate result
\begin{eqnarray}
{\cal A}^{(1,0)}  \left( \, \OM \, ; \HZB  \, \right)
& = &  \left\{ \,
\pab \, \left[ \, \frac{ \delta Z + \TH \, (\delta \TH )}{(D\TH )^2} \
{\cal S} \, \right]
\ + \ 2 \
\frac{ \pab Z + \TH \, (\pab \TH )}{(D\TH )^2} \
\frac{\delta D \TH}{D \TH} \
{\cal S}  \right.
\\
& &  \left.
\ \ \ \ \ \ \ \ \
-  \
\frac{ (\delta \TH )( \pab \TH )}{(D\TH )^2} \
{\cal S} \, \right\}
\ - \
( \ \delta \leftrightarrow \pab \ )
\nonumber
\ \ \ ,
\end{eqnarray}
one finds that
\begin{equation}
2\, \Gamma_{WZ}^{(1,0)}  [ \varphi \, ; \HZB ] \ = \
\int_{{\bf SC}} d^3 z \ \left\{ \, (\HZB )^{\varphi}  \ \pa D \, {\rm ln} \,
\LA^{\varphi} \ - \
\HZB \ \pa D \, {\rm ln} \,
\LA \, \right\}
\ \ \ .
\end{equation}
For $(\HZB )^{\varphi} = 0$, this expression
reduces to the previously considered
WZP functional \cite{dg1}.

\vskip 1.5truecm

{\bf \Large{Acknowledgements}}

\vspace {5mm}

I am grateful to S.Lazzarini for communicating reference \cite{sl}
and to R.Stora for some discussions on the
bosonic theory.
My special thanks go to F.Delduc for his interest in this work and
for his assistance in deriving
eq.(\ref{305}).
Also, I wish to thank
D.Z.Freedman, J.Ellis and P.Breitenlohner for the hospitality
extended to me at
MIT, CERN and MPI at different stages of this work.


\newpage

\begin{thebibliography}{22}
\newcommand{\artref}[4]{{\sc #1}, {\it #2} {\bf #3} #4}
\newcommand{\bookref}[2]{{\sc #1}, #2}




\bibitem{kn}
\bookref
{S.Kobayashi and K.Nomizu}{``Foundations of Differential Geometry I, II",
(John Wiley and Sons, New York 1963, 1969) ;}

\bookref
{D.H.Sattinger and O.L.Weaver}{``Lie groups and algebras with applications to
Physics, Geometry and Mechanics", (Springer-Verlag, New York 1986) .}



\bibitem{tr}
\bookref
{T.Regge}{in ``Relativity, Groups and Topology II", Les Houches,
Session XL, 1983,
B.S.De Witt and R.Stora, eds. (North Holland, 1984) .}



\bibitem{jm}
\bookref
{J.Milnor}{in ``Relativity, Groups and Topology II", Les Houches,
Session XL, 1983,
B.S.De Witt and R.Stora, eds. (North Holland, 1984) .}





\bibitem{sl}
\bookref
{S.Lazzarini}{Th\`ese de doctorat ``Sur les mod\`eles conformes
Lagrangiens bidimensionnels",
LAPP/Annecy, April 1990 .}



\bibitem{klt}
\artref
{M.Knecht, S.Lazzarini and F.Thuillier}{Phys.Lett.}{B251}{(1990) 279 .}






\bibitem{nr}
\artref
{Y.Ne'eman and T.Regge}{Riv.Nuov.Cim.}{5}{(1978) 1 ;}

\bookref
{F.Gieres}{``Geometry of supersymmetric gauge theories",
Lecture Notes in Physics 302 (Springer Verlag, 1988) .}







\bibitem{dg}
\artref
{F.Delduc and F.Gieres}{Class. Quantum Grav.}{7}{(1990) 1907 .}





\bibitem{dg1}
\artref
{F.Delduc and F.Gieres}{Int.J.Mod.Phys.}{A7}{(1992) 1685 .}







\bibitem{rs}
\bookref
{R.Stora}{in ``New perspectives in Quantum Field Theory", J.Abat,
M.Asorey and A.Cruz, eds. (World Scientific, Singapore 1986) .}










\bibitem{df}
\bookref
{D.Friedan}{in ``Unified String Theories", Santa Barbara Workshop,
M.B.Green and D.Gross, eds. (World Scientific, Singapore 1986) .}



\bibitem{gn}
\artref
{J.Grundberg and R.Nakayama}{Mod.Phys.Lett.}{A4}{ (1989) 55 .}


\bibitem{gg}
\artref
{S.J.Gates, Jr. and F.Gieres}{Nucl.Phys.}{B320}{(1989) 310.}

\end{thebibliography}
\end{document}